\documentclass{amsart}
\usepackage[utf8]{inputenc}
\usepackage{graphicx}
\usepackage{float}
\usepackage[foot]{amsaddr}

\title[Estimates of COVID-19 infected individuals]{Estimates of the proportion of SARS-CoV-2 infected individuals in Sweden}

\author{Henrik Hult} 
\author{Martina Favero}
\address{Department of Mathematics, KTH Royal Institute of Technology}
\date{May 21, 2020}
\email{hult@kth.se; mfavero@kth.se}

\begin{document}

\maketitle

\begin{abstract}
    In this paper a Bayesian SEIR model is studied to estimate the proportion of the population infected with  SARS-CoV-2, the virus responsible for  COVID-19. To capture heterogeneity in the population and the effect of interventions to reduce the rate of epidemic spread, the model uses a time-varying contact rate, whose logarithm has a Gaussian process prior. A Poisson point process is used to model the occurrence of deaths due to COVID-19 and the model is calibrated using data of daily death counts in combination with a snapshot of the the proportion of individuals with an active infection, performed in Stockholm in late March. The methodology is applied to regions in Sweden. The results show that the estimated proportion of the population who has been infected is around $13.5\%$ in Stockholm, by 2020-05-15,  and ranges between $2.5\% - 15.6\%$ in the other investigated regions. In Stockholm  where the peak of daily death counts is likely behind us, parameter uncertainty does not heavily influence the expected daily number of deaths, nor the expected cumulative number of deaths. It does, however, impact the estimated cumulative number of infected individuals. In the other regions, where random sampling of the number of active infections is not available, parameter sharing is used to improve estimates, but the parameter uncertainty remains substantial. 
\end{abstract}



\section{Introduction}

To understand the spread of the novel coronavirus, SARS-CoV-2, at an aggregate level it is possible to model the dynamic evolution of the epidemic using standard epidemic models. Such models include the (stochastic) Reed-Frost model and more general Markov chain models, or the corresponding  (deterministic) law of large numbers limits such as the general epidemic model, see \cite{Britton20}. 
There is an extensive literature on extensions of the standard epidemic models incorporating various degrees of heterogeneity in the population, e.g.~age groups, demographic information, spatial dependence, etc. These additional characteristics make the models more realistic. For instance, it is possible to evaluate the effect of various intervention strategies. More complex models also involve additional parameters that need to be estimated, contributing to a higher degree of parameter uncertainty.  

A problem when calibrating, even the standard epidemic models, to COVID-19 data is that there are few reliable sources on the number of infected individuals. Publicly available sources provide data on the number of positive tests, the number of hospitalizations, the number of ICU admission and the number of deaths due to COVID-19. In some cases, small random samples of an active infection may be available. For example, the Swedish Folkh\"alsomyndigheten performed such a test in Stockholm with about 700 subjects in early April 2020. Moreover, there is still no consensus in the literature on the value of important parameters such as the basic reproduction number $R_0$ and the infection fatality rate. 

A useful approach to incorporate the parameter uncertainty in the models is to consider a Bayesian framework. In the Bayesian approach parameter uncertainty is quantified by prior distributions over the unknown parameters.  The impact of observed data, in the form of a likelihood, yields, via Bayes' theorem, the posterior distribution, which quantifies the effects of parameter uncertainty. The posterior can be used to construct estimates on the number of infected individuals, predictions on the future occurrence of infections and deaths, as well as uncertainties in such estimates. 

In this paper an SEIR epidemic model with time-varying contact rate will be used to model the evolution of the number of susceptible (S), exposed (E), infected (I), and recovered (R) individuals. A time varying contact rate is used to capture heterogeneity in the population, which causes the rate of the spread of the epidemic to vary as the virus spreads through the population. Moreover, the time varying contact rate allows modeling the effect of interventions aimed at reducing the rate of epidemic spread.  A Poisson point process is introduced to model the occurrence and time of deaths. Random samples of tests for active infections are treated as binomial trials where the success probability is the proportion of the population in the infectious state. 

The methods are illustrated on regional data of daily COVID-19 deaths in Sweden. It is demonstrated that, by combining the information in the observed number of deaths and random samples of active infections, fairly precise estimates on the number of infected individuals can be given. By assuming that some parameters are identical in several regions,  estimates for regions outside Stockholm can also be provided, albeit with greater uncertainty. 

Our approach is inspired by \cite{Chatzilena19} where the authors considers a Bayesian approach to model an influenza outbreak. The main extensions include the introduction of the Poisson point process to model the occurrence of deaths, the addition of random sampling to test for infection, and an extension to multiple regions. To evaluate the posterior distribution we employ Markov chain Monte Carlo (MCMC) sampling. Samples from the posterior are obtained using the Hamiltonian Monte Carlo algorithm, NUTS, by \cite{Hoffmann14}, implemented in the software Stan, which is an open source software for MCMC.  

\section{Background} \label{sec:background}
To model the spread of the epidemic we consider the deterministic SEIR model \cite{Anderson92,Diekmann13}, which is a simple deterministic model describing the evolution of the number of susceptible, exposed, infected, and recovered individuals in a large homogeneous population with $N$ individuals. The epidemic is modeled by $\{(S_t, E_t, I_t), t \geq 0\}$, where $S_t, E_t$ and $I_t$ represent the number of susceptible, exposed  and infected individuals at time $t$, respectively. The total number of recovered and deceased individuals at time $t \geq 0$ is always given by $N - S_t - E_t- I_t$. The epidemic starts from a state $S_0, E_0,I_0$ with $S_0 + E_0 + I_0 = N$, and proceeds by updating, 
\begin{align*}
    S_{t+1} &= S_t - \beta \frac{S_t I_t}{N}, \\
    E_{t+1} &= E_t + \beta \frac{S_t I_t}{N} - \nu E_t, \\
    I_{t+1} &= I_t + \nu E_t - \gamma I_t. 
\end{align*}
The parameters are the contact rate $\beta > 0$, the rate $\nu$ of transition from the exposed to the infected state and the recovery rate $\gamma > 0$.  Note that $I_t$ represents the number of individuals with an active infection at time $t$, whereas $N-S_t$ is the cumulative number of individuals who have been exposed, and possibly infected, recovered or deceased, up until time $t$. 

In the context of the COVID-19 epidemic the contact rate cannot be assumed to be constant, primarily due to interventions implemented in the early stage of the epidemic. Moreover, as the SEIR model describes the evolution at an aggregate level, a time varying contact rate may be used to capture inhomogeneities in the population. If, for example, the epidemic is initiated in a rural area the contact rate may be rather low, but as the epidemic reaches major cities the contact rate will be higher. The resulting SEIR model with time varying contact rate is given by 
\begin{align}\label{eq:SEIR}
\left\{\begin{array}{l}
    S_{t+1} = S_t - \beta_t \frac{S_t I_t}{N}, \\
    E_{t+1} = E_t + \beta_t \frac{S_t I_t}{N} - \nu E_t, \\
    I_{t+1} = I_t + \nu E_t - \gamma I_t. 
    \end{array}, \right.
\end{align}
Clearly, one needs to put some restriction on the amount of variation of the contact rate. In this paper a Gaussian process prior will be used on the log contact rate, which restricts the amount of variation in time, but is sufficiently flexible to capture the reduction in contact rate after the interventions. 

When observations on the number of infected and recovered individuals are available, the model \eqref{eq:SEIR} can be fitted to these observations. In the context of COVID-19, observations on the number of infected and recovered individuals are unavailable. There are many symptomatic individuals who are not tested and potentially a large pool of asymptomatic individuals. In this paper we will rely on the number of registered deaths due to COVID-19 to calibrate the model. In addition we will incorporate the test results from a random sample that provides a snapshot on the number of individuals with an active infection.  


\section{A Poisson point process for the occurrence of deaths}

To model the occurrence of deaths due to COVID-19 we consider the following Poisson point process representation. We refer to \cite{Kallenberg17} for details on Poisson point processes.  Let $f$ denote the infection fatality rate, that is, the probability that an infected individual eventually dies from the infection. Consider the number of individuals that enters the infected state on day $t$, that is, $\nu E_t$. Each such infected individual has probability $f$ to eventually die from the infection. Conditional on death due to the infection, the time from infection until death is assumed independent of everything else and follows a probability distribution with probability mass function $p_{s_D}$. Each individual that dies may be represented as a point $(t, \tau)$ in $E := \{(t,\tau) \in \mathbb{N}^2: \tau \geq t\}$, where $t$ denotes the time of entry to the infected state and $\tau$ the time of death of the individual. The number of deaths at time $\tau$ can then be computed by counting the number of points on the line $\cup_{t=0}^\tau (t,\tau)$.   

The number of deaths, and the corresponding time of infection and time of death is conveniently modelled by a Poisson point process on $E$.  Let $\xi$ be a Poisson point process on $E$ with intensity 
\begin{align}
    \nu(t,\tau) = f \nu E_t p_{s_D}(\tau-t). 
\end{align}
We may interpret a point at $(t, \tau)$ of the Poisson point process as the time of infection, $t$, and the time of death, $\tau$, of an individual who dies from the infection. 

The number of deaths $D_\tau$ that occurs at time $\tau$ is then given by summing up all the points of the point process on the row corresponding to $\tau$,  $\xi(\cup_{t = 0}^\tau (t,\tau))$. Since the rows are disjoint this implies that $D_0, D_1, \dots$ are independent with each $D_\tau$ having a Poisson distribution with parameter 
\begin{align*}
    \lambda_\tau = \nu(\cup_{t = 0}^\tau (t,\tau)) = f \nu \sum_{t=0}^\tau  E_t p_{s_D}(\tau-t). 
\end{align*}
Throughout this paper $p_{s_D}$ is the probability mass function of a negative binomial distribution with mean $s_D$. More precisely, a parametrization of the negative binomial distribution with parameters $r, s_D$ will be used, where
\begin{align*}
    p_{s_D}(n) = \left(\frac{s_D}{r + s_D} \right)^n \left(\frac{r}{r+ s_D} \right)^r \binom{n+r-1}{r-1}.
\end{align*}
The value $r = 3$ will be used throughout as this fits well with the distribution of observed duration from symptoms to death in the study by \cite{Wu20}.  
\section{Prior distributions and derivation of the likelihood}
In this section we provide the assumptions on the prior distributions and derive the expression of the likelihood of the model.
Note that $\lambda_\tau$ is a function of all the parameters of the model, $\theta = (\{\beta_t\}, \nu, \gamma, s_0, f, s_D)$. The parameters, their interpretation and prior distribution are summarized in Table \ref{tab:priors}. Actually, since the contact rate is positive, a Gaussian process (GP) prior will be used for the natural logarithm of the contact rate, denoted log-GP in the sequel. The Gaussian process has a constant mean $\mu$ and a squared-exponential covariance kernel $k$ with parameters $\alpha, \rho, \delta$ such that 
\begin{align*}
    \text{Cov}(\log \beta_{t_1}, \log \beta_{t_2}) = k(t_2 - t_1) = \alpha^2 \exp\left(-\frac{1}{2\rho^2}|t_2-t_1|^2\right) + \delta^2 I\{t_2 = t_1\}.
\end{align*}
\begin{table}[!ht]
    \centering
    \begin{tabular}{c|l|l|l}
        Parameter & Explanation & Prior distribution & Hyper-parameters \\
        \hline
         $\{\beta_t\}$ & contact rate & log-GP$(\mu, k)$ & $\mu, \alpha, \rho, \delta$ \\ 
         $\nu$ & rate from exposed to infected & Gamma$(a_{\nu}, b_{\nu})$ &  $a_{\nu}, b_{\nu}$ \\
         $\gamma$ & recovery rate & Gamma$(a_{\gamma}, b_{\gamma})$ &  $a_{\gamma}, b_{\gamma}$ \\
         $s_0$ & initial susceptible fraction & Beta$(a_{s_0}, b_{s_0})$ & $a_{s_0}, b_{s_0}$ \\
         $f$ & infection fatality rate & Beta$(a_f, b_f)$ & $a_f, b_f$ \\
         $s_D$ & expected duration & Gamma$(a_{s_D}, b_{s_D})$ & $a_{s_D}, b_{s_D}$
    \end{tabular}
    \caption{Specification of the parameters and prior distributions. }
    \label{tab:priors}
\end{table}
To compute the likelihood the observed number of daily deaths, $d_0, d_1, \dots, d_T,$ will be used, in combination with a random sample of $n_0$ tests for active infection, performed at a time $t_0$, when such test result is available. The number of individuals $Z$ with positive test result has a Bin$(n_0, I_{t_0}/N)$ distribution. The full likelihood is given by: 
\begin{align}
\label{eq:likelihood}
    p_{D,Z \mid \Theta}(d_1, \dots, d_T, z \mid \theta) &= \prod_{\tau = 1}^T \text{Prob}(D_\tau = d_\tau)  \times \text{Prob}(Z = z) \nonumber\\ &=  \prod_{\tau = 1}^T \frac{\lambda_\tau^{d_\tau}}{d_\tau !} e^{-\lambda_\tau} \times \binom{n_0}{z}\left(\frac{I_{t_0}}{N}\right)^z\left(1-\frac{I_{t_0}}{N}\right)^{n_0 - z}. 
\end{align}
The joint prior is the product of the marginal priors and leads, by Bayes's theorem, to the posterior, 
\begin{align*}
    p_{\Theta \mid D, Z}(\theta \mid d_1, \dots, d_T, z) \propto p_{D, Z \mid \Theta}(d_1, \dots, d_T, z \mid \theta) p_\Theta(\theta).
\end{align*}

The expected number of daily deaths $\lambda_\tau$, the cumulative number of deaths and the cumulative number of infected individuals $ N - S_t$ are all functions of $\theta$ and their distribution can therefore be inferred from the posterior
$p_{\Theta \mid D, Z}$. By sampling from $p_{\Theta \mid D}$ and iterating the dynamics \eqref{eq:SEIR} estimates of these quantities may be obtained along with the effects of parameter uncertainty. Moreover, predictions on the future development of the above mentioned quantities can be obtained by extrapolating the contact rate into the future. 

As the posterior distribution is unavailable in explicit form it is necessary to employ Monte Carlo methods. In the next section Markov chain Monte Carlo methods are briefly described to sample from the posterior. 

\subsection{Multiple regions}
The SEIR model \eqref{eq:SEIR} and the derivation of the likelihood \eqref{eq:likelihood} considers a single region. In the context of multiple regions it may be reasonable to assume that some parameters are identical. For example, when considering multiple regions of Sweden below it will be assumed that the rate, $\nu$, from exposed to infected, the recovery rate, $\gamma$, the infection fatality rate, $f$, and the duration, $s_D$ are identical in all regions. It is tempting to include interaction terms between the regions as infected individuals from one region may travel to another region and cause new infections. In this paper, it will be assumed that each region has its own time varying contact rate that incorporates fluctuations in new infections due to import cases from other regions. 

The likelihood from multiple regions is simply the product of the marginal likelihood for the individual regions and the prior is the product of the marginal priors for each parameter. Thus, for two regions the prior will be the product of two Gaussian process priors for the respective log contact rates for the two regions and the product of the marginal priors for the remaining parameters.

\section{Markov chain Monte Carlo}

Markov chain Monte Carlo (MCMC) methods in Bayesian analysis aims at sampling from the posterior distribution. This is non-trivial because the marginal distribution of the data, which acts as normalizing constant of the posterior is practically impossible to compute. In MCMC algorithms the posterior is represented as a target distribution. The algorithms rely on the construction of a Markov chain whose invariant distribution is the target distribution. Standard MCMC methods are based on acceptance-rejection steps, where random proposals are accepted or rejected with a probability that does not require knowledge of the normalizing constant, e.g., Metropolis-Hasting and Gibbs sampling \cite{Metropolis53, Hastings70, Geman84}. When the target distribution is complex and multi-modal, standard methods may lead to poor mixing of the Markov chain and slow convergence to the target distribution. 

To overcome slow mixing of the Markov chain gradient-based sampling can be applied, which adapt the proposal distribution based on gradients of the target, see e.g.~ \cite{Betancourt17}. In this paper we will employ a Hamiltonian Monte Carlo sampler, the No-U-turn sampler (NUTS) by \cite{Hoffmann14} in combination with automatic
differentiation to numerically approximate the gradients
\cite{Griewank08}, which is implemented in the open source software Stan.  

\section{Estimates and predictions for regions of Sweden}
In this section the estimates of the number of infected individuals and predictions on the evolution of the number of deaths and number of infected individuals are provided for ten regions of Sweden. The epidemic is considered to start on 2020-03-01 and interventions in Sweden began on 2020-03-16. The joint prior distribution is the product of the marginal priors, and the hyper-parameters are specified in Table \ref{tab:STO_priors}. The choices of hyper-parameter values are made in line with existing literature on the COVID-19 epidemic. As a general principle we have used informative priors on the parameters $\nu, \gamma,$ and $s_D$, whereas the priors on the time-varying contact rate $\{\beta_t\}$ and the fatality rate $f$ are uninformative. Folkh\"alsomyndigheten reports that the incubation period is usually around $5$ days, which corresponds to $1/\nu \approx 5$. Similarly the expected time to recovery is around $14$ days, $1/\gamma \approx 14$. The overall infection fatality rate $f$ is estimated to be in the range 
$0.003-0.013$, see \cite{Verity20, Wu20}. However, since the infection fatality rate is a very important parameter we have used an uninformative prior and simply use a uniform prior, Beta$(1,1)$. The expected duration from symptoms to death is around 16 days, see \cite{Wu20}.  

Samples from the posterior are obtained using the NUTS-sampler
with a burn-in period of $500$ samples and $5\,000$ samples after burn-in.  

\begin{table}[!ht]
    \centering
    \begin{tabular}{c|l|l|r|r}
        Parameter & Prior distribution & Hyper-parameters & Mean & Prior $95\%$-C.I.\\
        \hline
         $\{\beta_t\}$ & log-GP$(\mu, k)$ & $\mu = 0 ,\; \alpha = 1,$   & $1.13$ & $[0.38, 2.66]$  \\ 
         & & $\rho = 15,\; \delta = 10^{-3}$ & & \\
         $\nu$  & Gamma$(a_{\nu}, b_{\nu})$ &  $a_\nu = 100, b_\nu = 500$& $0.2$ & $[0.16, 0.24]$ \\
         $\gamma$  & Gamma$(a_{\gamma}, b_{\gamma})$ &  $a_\gamma = 100, b_\gamma = 1400$& $0.071$ & $[0.058, 0.086]$ \\
         $s_0$ & Beta$(a_{s_0}, b_{s_0})$ & $a_{s_0} = 100, b_{s_0} = 0.5$ & $0.99$ & $[0.97, 1.0]$ \\
         $f$  & Beta$(a_f, b_f)$ & $a_f = 1, b_f = 1$ & $0.5$ & $[0.025, 0.975]$ \\
         $s_D$  & Gamma$(a_{s_D}, b_{s_D})$ & $a_{s_D}= 900, b_{s_D} = 50$ & 18.0& $[16.8, 19.2]$ \\
         \hline
    \end{tabular}
    \caption{Prior distributions, hyper-parameters and prior credibility intervals.}
    \label{tab:STO_priors}
\end{table}

\subsection{Region Stockholm}
The Region Stockholm has $N = 2.34$ million inhabitants. The daily death counts in Region Stockholm from 2020-03-01 -- 2020-05-15 are obtained from the webpage: \verb=https://c19.se/=. On 2020-04-09 the Swedish authority, Folkh\"alsomyndigheten, published the results 
of a random sample of $707$ individuals performed between 2020-03-27 and 2020-04-03\footnote{https://www.folkhalsomyndigheten.se/nyheter-och-press/nyhetsarkiv/2020/april/resultat-fran-undersokning-av-forekomsten-av-covid-19-i-region-stockholm/}. It showed that 18 individuals carried the SARS-CoV-2 virus. These results are included in the analysis as a binomial sample of size $n_0 = 707$ and success probability $I_{t_0}/N$ where the test date, $t_0$, is assumed to be 2020-03-30. 

\begin{figure}[!ht]
    \centering
    \includegraphics[scale=0.5]{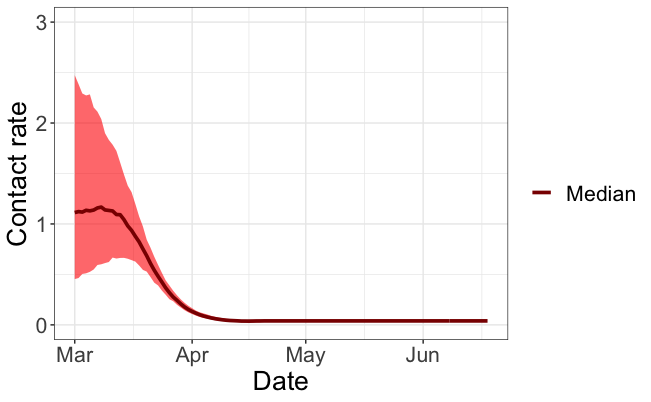}
    \caption{Estimated contact rate for Region Stockholm until 2020-05-01 based on data from 2020-03-01 -- 2020-05-15. The graph shows the posterior median and point-wise $95\%$ credibility interval. After 2020-05-01 the contact rate is extrapolated, by assuming it will remain constant.}
    \label{fig:Contact_rate_STO}
\end{figure}

A summary of the marginal posterior distributions is provided in Table \ref{tab:STO_posts}. The posterior distribution of the time varying contact rate is illustrated in Figure \ref{fig:Contact_rate_STO}. Note that although there is great uncertainty about the initial contact rate, the model clearly picks up the reduction in contact rate after the interventions began on 2020-03-16. The contact rate is gradually reduced around the time of intervention and then remains at a low level. This slow reduction of the contact is, however, not due to stiffness of the Gaussian process kernel. We have experimented with a sharp break-point in the contact rate at the time of intervention, but it did not provide more accurate results. On the contrary, the data suggests that the reduction of the contact rate is slow. The contact rate is estimated until 2020-05-01. After this date the posterior is unreliable. This is because many of the deaths of individuals who are infected after 2020-05-01 have not yet been observed. For this reason, the contact rate is only estimated until 2020-05-01. 

To perform estimates and predictions on the future number of daily and cumulative infections and deaths, the contact rate has been extrapolated from its value on 2020-05-01. The posterior distribution suggests that the contact rate is constant, at a low rate, since roughly 2020-04-07, which motivates extrapolation into the future, assuming that the interventions remains at the present level.  

\begin{figure}[!ht]
    \centering
    \includegraphics[scale=0.25]{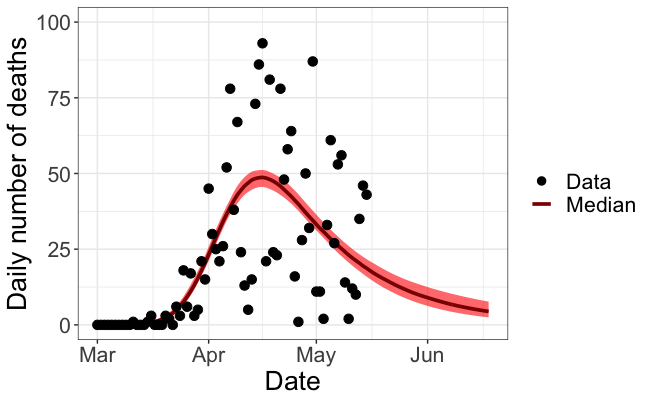}
    \includegraphics[scale=0.25]{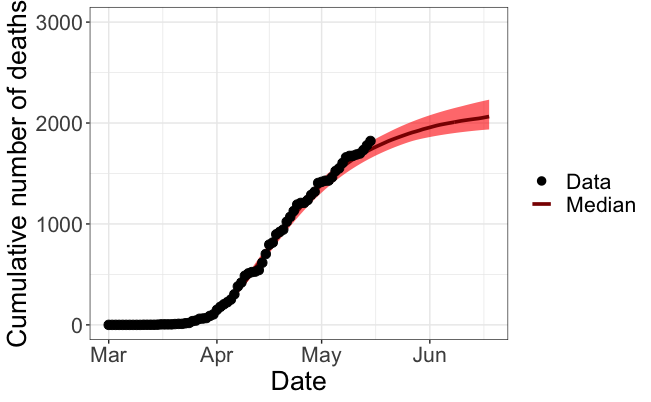}\\
    \includegraphics[scale=0.25]{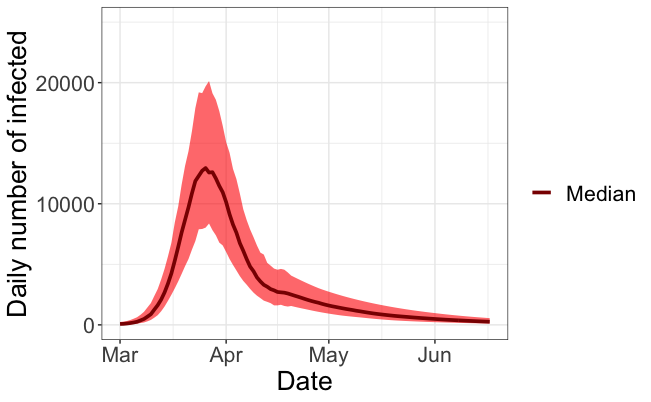}
    \includegraphics[scale=0.25]{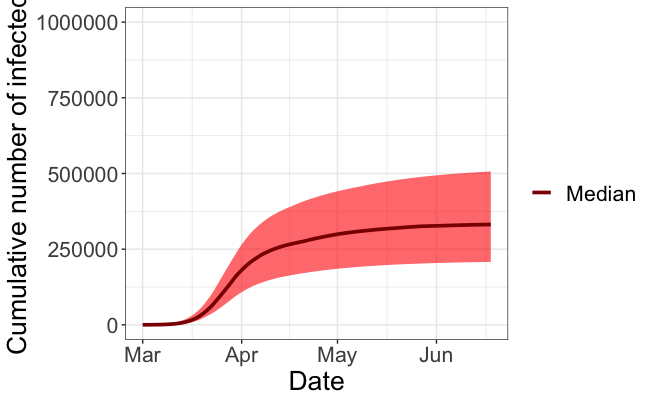}
    \caption{Data from Region Stockholm until 2020-05-16. \emph{Top left:} Observed daily number of deaths (black dots), the posterior median (dark red) and $95\%$ credibility interval for the expected daily number of deaths. \emph{Top right:} Observed cumulative number of deaths (black dots), the posterior median (dark red) and $95\%$ credibility interval for the expected cumulative number of deaths. \emph{Bottom left:} The posterior median (dark red) and $95\%$ credibility interval for the daily number of infected individuals. \emph{Bottom right:} The posterior median (dark red) and $95\%$ credibility interval for the cumulative number of infected individuals. }
    \label{fig:Daily_deaths_STO}
\end{figure}

Figure \ref{fig:Daily_deaths_STO} (top left) shows the observed daily number of deaths (black dots) along with the posterior median (dark red) and $95\%$  credibility interval (red) for the expected number of daily deaths. Figure \ref{fig:Daily_deaths_STO}  (top right) shows the observed cumulative number of deaths (black dots) along with the posterior median (dark red) and $95\%$  credibility interval (red) for the expected cumulative number of deaths. 

We observe that the parameter uncertainty does not substantially impact the expected number of daily deaths and the peak of the daily number of deaths appears to have occurred by mid April. Similarly, the expected cumulative number of deaths in Stockholm is likely to terminate slightly above 2000. 

We emphasize that this is the \emph{expected number of deaths}, $\lambda_\tau$. Since we are considering a Poisson distribution for the number of daily deaths an approximate $95\%$-prediction interval would be $\lambda_\tau \pm 2 \sqrt{\lambda_\tau}$, where $\lambda_\tau$ is the Poisson parameter on day $\tau$. 

Note from the observed number of daily deaths that the empirical distribution of daily deaths appear to be overdispersed, the variance is substantially larger than the mean. This is likely due to reporting of the data. The data presented at \verb=https://c19.se/= does not correct the reporting of death dates in hindsight. A comparison at the national level with data provided by Folkh\"alsomyndigheten shows that the official records of the daily number of deaths for Sweden does not appear to be overdispersed. Nevertheless, even after smoothing the data from \verb=https://c19.se/= by a moving average over a few days, the results of the simulations remain essentially the same. 

Figure \ref{fig:Daily_deaths_STO}  (bottom left/right) shows the posterior median (dark red) and $95\%$  credibility interval (red) for the daily/cumulative number of infected individuals. Although the parameter uncertainty has significant impact on the cumulative number of infected individuals, some conclusions are still possible. As of mid May, the cumulative number of infected individuals has almost reached its terminal value and the spread of the epidemic has slowed down significantly. The estimated cumulative number of infected individuals is $13.5\%$ of the population in Stockholm. The estimated number of infected individuals by 2020-04-11 is $10.7\%$, showing  that these results are well in line with the reports of the anti-body test performed at KTH\footnote{https://www.kth.se/en/aktuellt/nyheter/10-procent-av-stockholmarna-smittade-1.980727}, which indicated that $10\%$ of the population in Stockholm had developed anti-bodies against the SARS-CoV-2 virus by the first weeks of April.  

We emphasize that the estimate of the cumulative number of infected individuals in Stockholm relies heavily on the inclusion of results from the random sampling performed by Folkh\"alsomyndigheten in late March, early April. Without this crucial piece of  information similar models to the one analyzed here may provide a significantly higher estimate on the cumulative number of infected.

\begin{table}[!ht]
    \centering    
    \begin{tabular}{c|r|r|r}
        Parameter & Post. mean & Post. $95\%$-C.I. & Prior $95\%$-C.I.\\
        \hline
         $\nu$ & $0.21$ & $[0.18, .25]$ & $[0.18, 0.24]$  \\ 
         $\gamma$  & $0.08$ & $[0.07, 0.10]$ & $[0.058, 0.086]$ \\
         $s_0$ &  $1.0$ & $[0.9997, 1.0]$ &  $[0.97, 1.0]$ \\
         $f$  & $0.007$ & $[0.004, 0.011]$ &  $[0.025, 0.0975]$ \\
         $s_D$  & $17.8$& $[16.7, 18.9]$ &  $[16.8, 19.2]$ \\
            \hline
    \end{tabular}
    \caption{Marginal posterior median and credibility intervals for Region Stockholm.}
    \label{tab:STO_posts}
\end{table}

\newpage
\subsection{Summary of the results for ten regions of Sweden}

In this section estimates of the cumulative number of infected individuals are provided for the following regions of Sweden: 
\begin{enumerate}
    \item Stockholm (population: $2.34 \cdot 10^6$) 
    \item V\"astra G\"otaland (population: $1.71 \cdot 10^6$) 
    \item \"Osterg\"otland (population: $1.36 \cdot 10^6$)
    \item \"Orebro (population: $3.02 \cdot 10^5$)
    \item Sk{\aa}ne (population: $1.36 \cdot 10^6$)
    \item J\"onk\"oping (population: $3.56 \cdot 10^5$)
    \item S\"ormland (population: $2.95 \cdot 10^5$)
    \item V\"astmanland (population: $2.74 \cdot 10^5$)
    \item Uppsala (population: $3.76 \cdot 10^5$)
    \item Dalarna (population: $2.87 \cdot 10^5$)
\end{enumerate}

\begin{figure}[!ht]
    \centering
    \includegraphics[scale=0.25]{Contact_rate_STO.png}
     \includegraphics[scale=0.46]{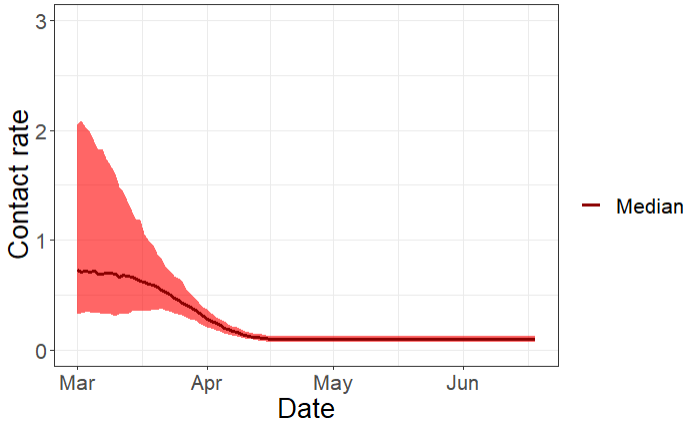}
    \includegraphics[scale=0.25]{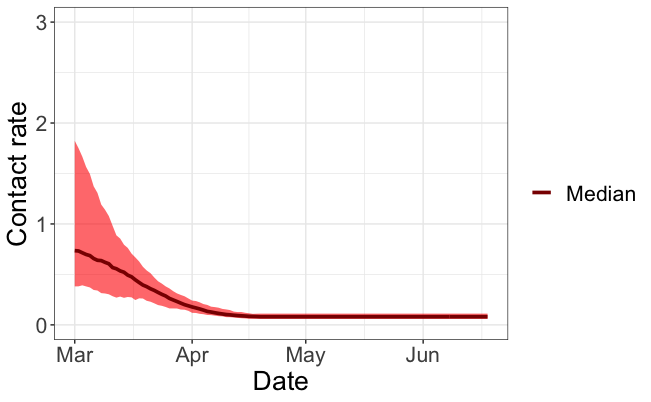}
    \includegraphics[scale=0.25]{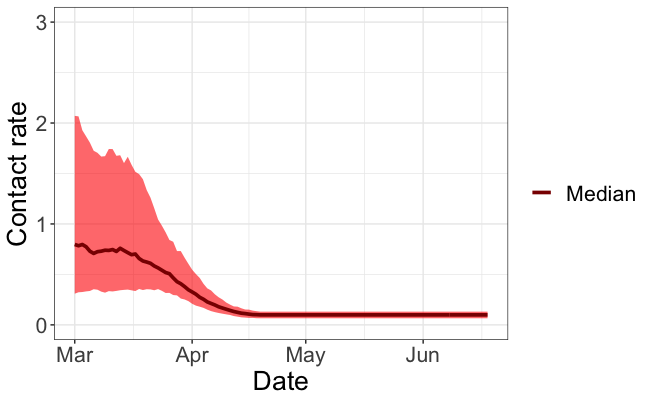}
    \includegraphics[scale=0.25]{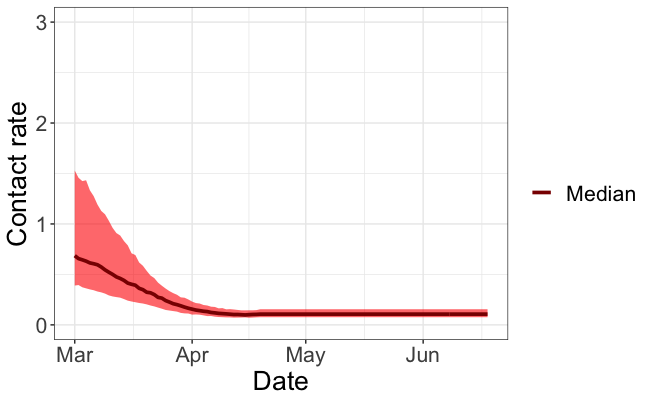}
    \includegraphics[scale=0.25]{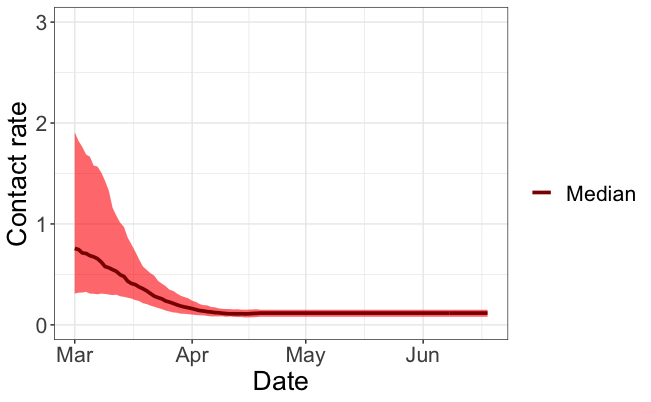}
    \includegraphics[scale=0.25]{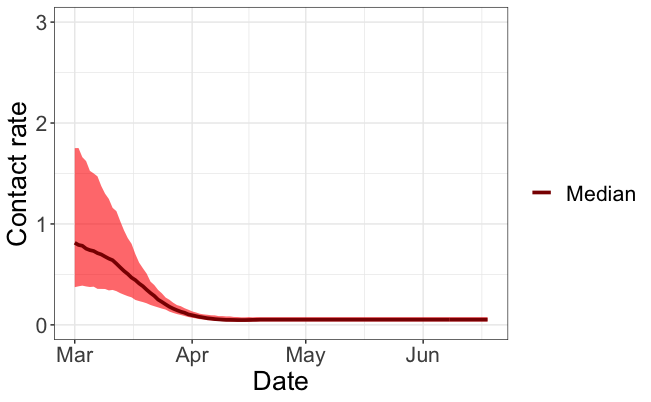}
    \includegraphics[scale=0.25]{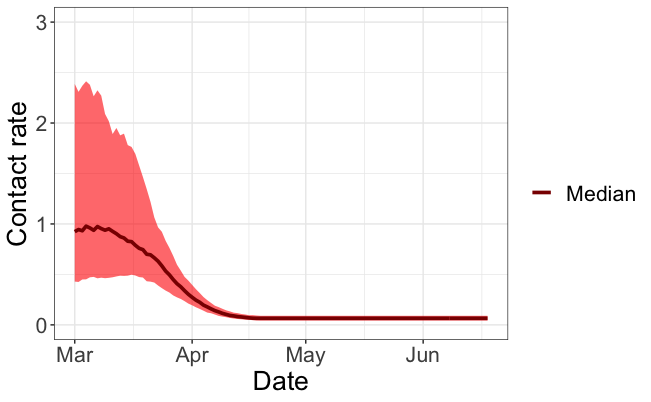}
    \includegraphics[scale=0.25]{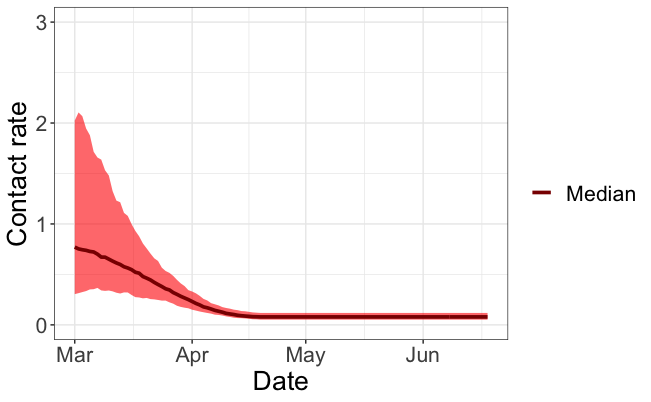}
    \includegraphics[scale=0.25]{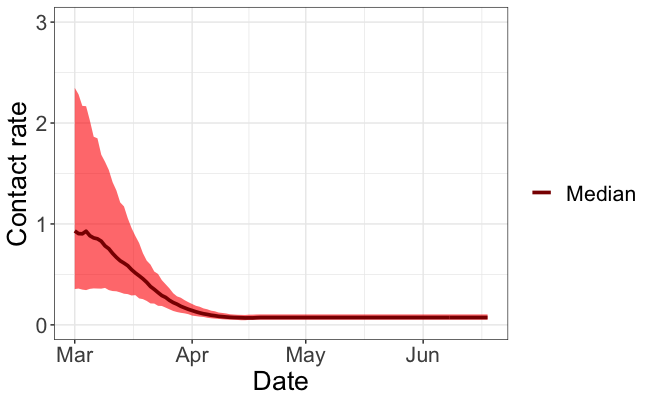}   
    \caption{Estimated contact rate for regions of Sweden until 2020-05-01 based on data from 2020-03-01 -- 2020-05-15. The graph shows the posterior median and point-wise $95\%$ credibility interval. After 2020-05-01 the contact rate is extrapolated, by assuming it will remain constant. From right to left, first row: Stockholm, V\"astra G\"otaland, second row:  \"Osterg\"otland, \"Orebro, third row: Sk{\aa}ne,  J\"onk\"oping, fourth row S\"ormland, V\"astmanland, fifth row: Uppsala, 
    Dalarna.}
    \label{fig:Contact_rates}
\end{figure}

\begin{figure}
    \centering
    \includegraphics[scale=0.46]{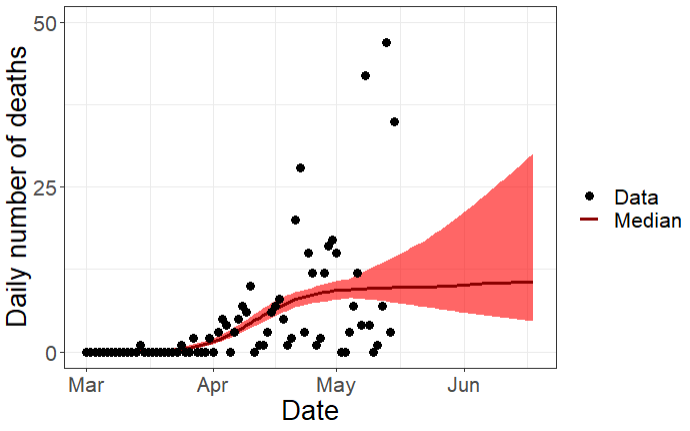}
    \includegraphics[scale=0.46]{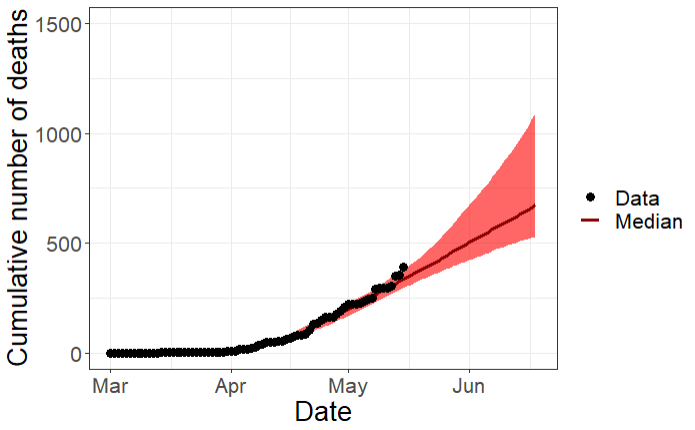}\\
    \includegraphics[scale=0.46]{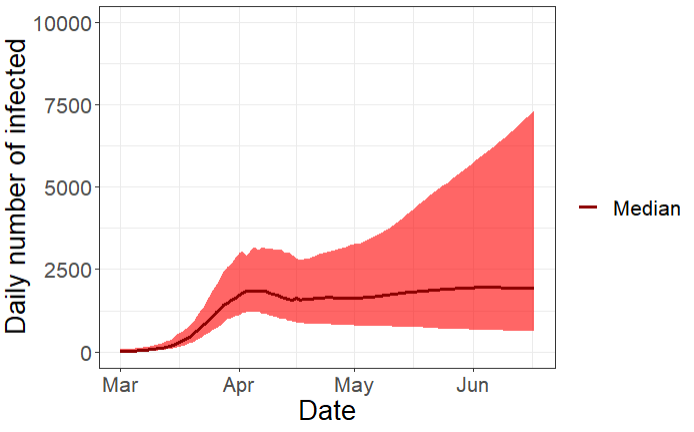}
    \includegraphics[scale=0.46]{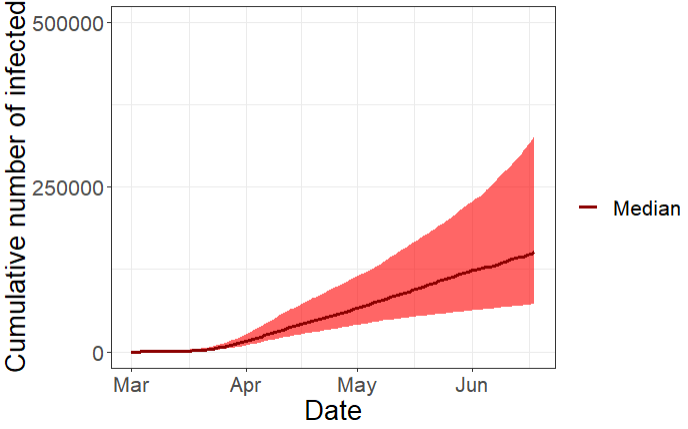}
    \caption{Data from Region V\"astra G\"otaland  until 2020-05-15. \emph{Top left:} Observed daily number of deaths (black dots), the posterior median (dark red) and $95\%$ credibility interval for the expected daily number of deaths. \emph{Top right:} Observed cumulative number of deaths (black dots), the posterior median (dark red) and $95\%$ credibility interval for the expected cumulative number of deaths. \emph{Bottom left:} The posterior median (dark red) and $95\%$ credibility interval for the daily number of infected individuals. \emph{Bottom right:} The posterior median (dark red) and $95\%$ credibility interval for the cumulative number of infected individuals. }
    \label{fig:Daily_deaths_VGO}
\end{figure}

\begin{figure}
    \centering
    \includegraphics[scale=0.25]{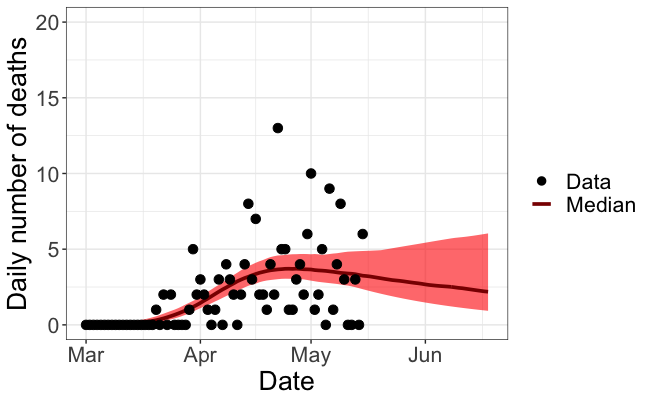}
    \includegraphics[scale=0.25]{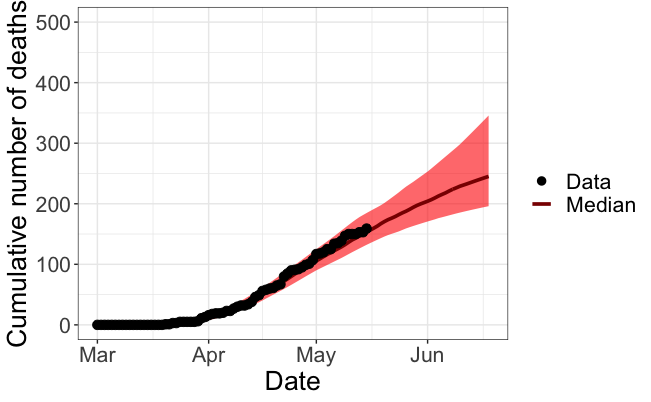}\\
    \includegraphics[scale=0.25]{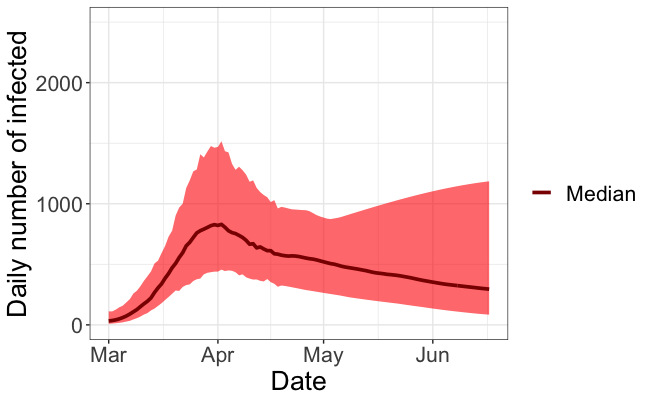}
    \includegraphics[scale=0.25]{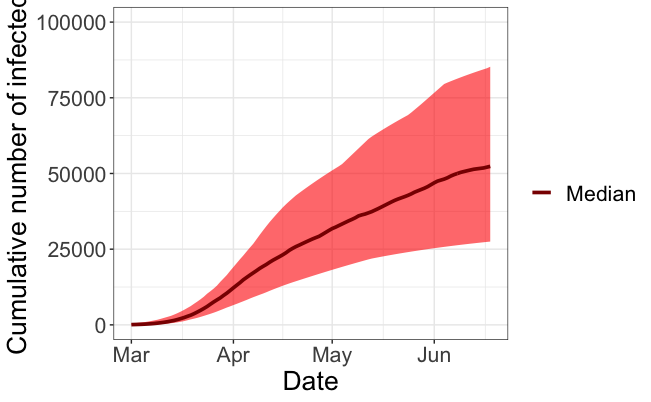}
    \caption{Data from Region \"Osterg\"otland  until 2020-05-15. \emph{Top left:} Observed daily number of deaths (black dots), the posterior median (dark red) and $95\%$ credibility interval for the expected daily number of deaths. \emph{Top right:} Observed cumulative number of deaths (black dots), the posterior median (dark red) and $95\%$ credibility interval for the expected cumulative number of deaths. \emph{Bottom left:} The posterior median (dark red) and $95\%$ credibility interval for the daily number of infected individuals. \emph{Bottom right:} The posterior median (dark red) and $95\%$ credibility interval for the cumulative number of infected individuals. }
    \label{fig:Daily_deaths_OGO}
\end{figure}

\begin{figure}
    \centering
    \includegraphics[scale=0.25]{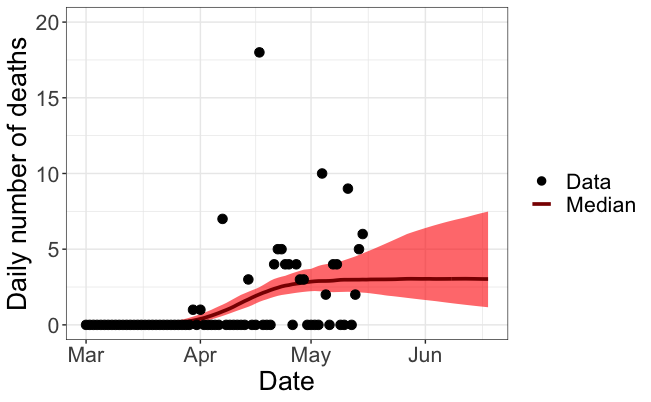}
    \includegraphics[scale=0.25]{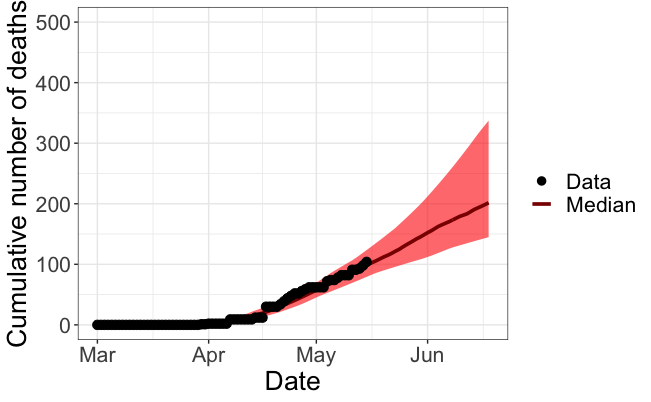}\\
    \includegraphics[scale=0.25]{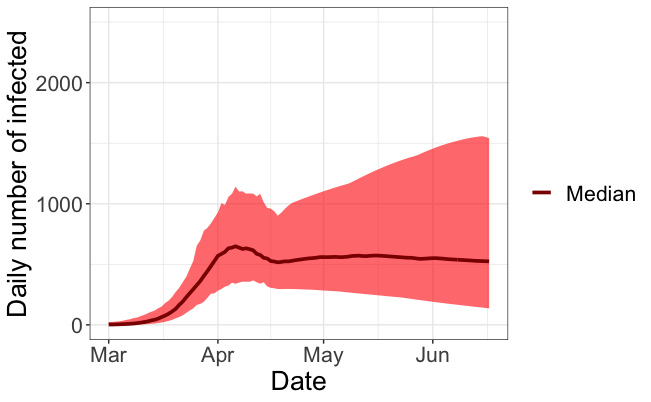}
    \includegraphics[scale=0.25]{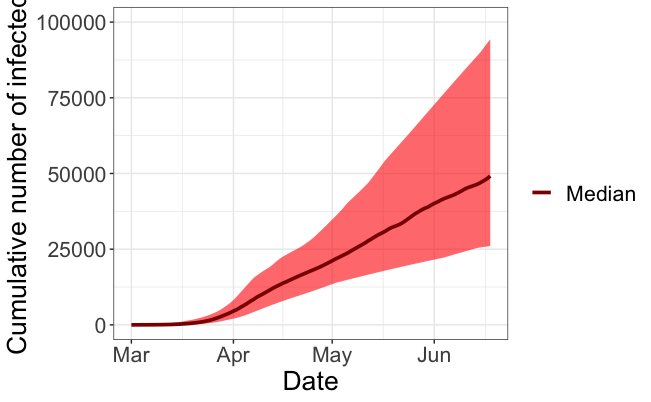}
    \caption{Data from Region \"Orebro  until 2020-05-15. \emph{Top left:} Observed daily number of deaths (black dots), the posterior median (dark red) and $95\%$ credibility interval for the expected daily number of deaths. \emph{Top right:} Observed cumulative number of deaths (black dots), the posterior median (dark red) and $95\%$ credibility interval for the expected cumulative number of deaths. \emph{Bottom left:} The posterior median (dark red) and $95\%$ credibility interval for the daily number of infected individuals. \emph{Bottom right:} The posterior median (dark red) and $95\%$ credibility interval for the cumulative number of infected individuals. }
    \label{fig:Daily_deaths_ORE}
\end{figure}

\begin{figure}
    \centering
    \includegraphics[scale=0.25]{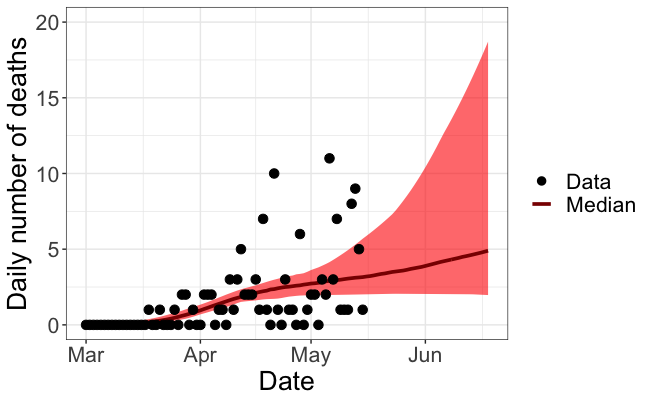}
    \includegraphics[scale=0.25]{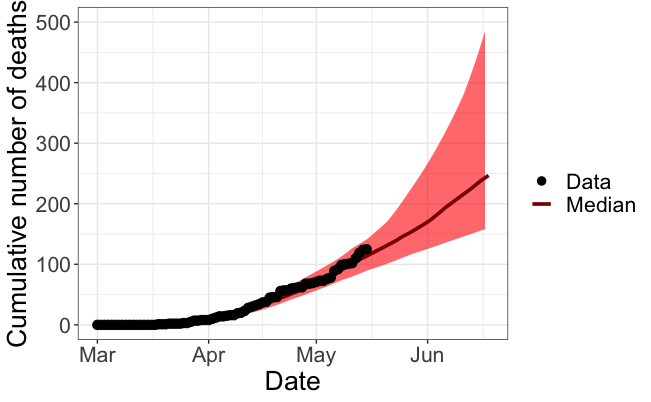}\\
    \includegraphics[scale=0.25]{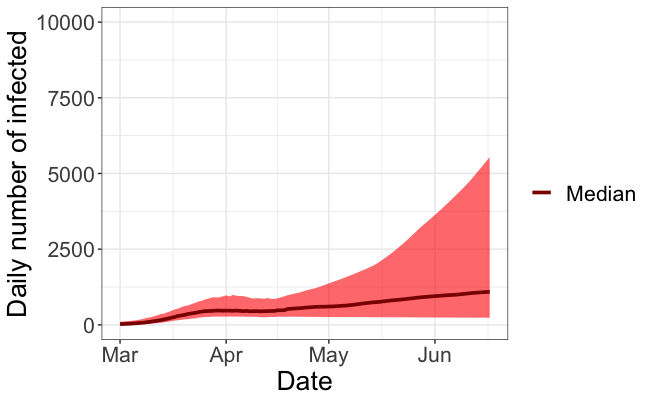}
    \includegraphics[scale=0.25]{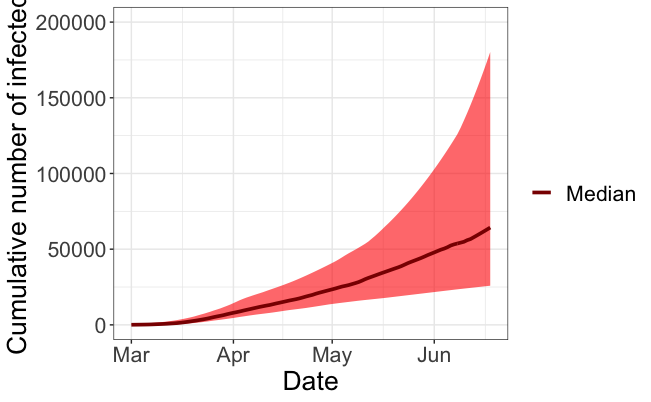}
    \caption{Data from Region Sk{\aa}ne  until 2020-05-15. \emph{Top left:} Observed daily number of deaths (black dots), the posterior median (dark red) and $95\%$ credibility interval for the expected daily number of deaths. \emph{Top right:} Observed cumulative number of deaths (black dots), the posterior median (dark red) and $95\%$ credibility interval for the expected cumulative number of deaths. \emph{Bottom left:} The posterior median (dark red) and $95\%$ credibility interval for the daily number of infected individuals. \emph{Bottom right:} The posterior median (dark red) and $95\%$ credibility interval for the cumulative number of infected individuals. }
    \label{fig:Daily_deaths_SKA}
\end{figure}

\begin{figure}
    \centering
    \includegraphics[scale=0.25]{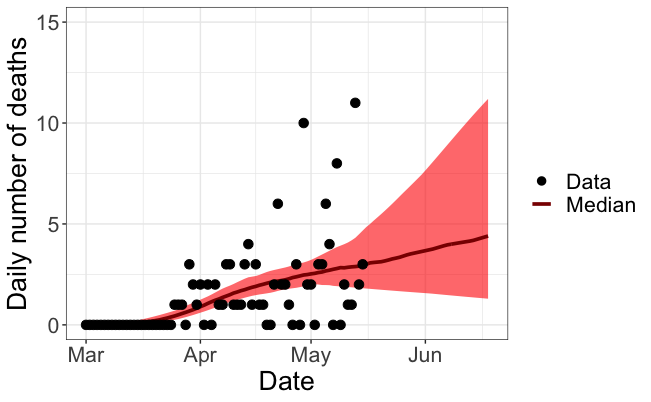}
    \includegraphics[scale=0.25]{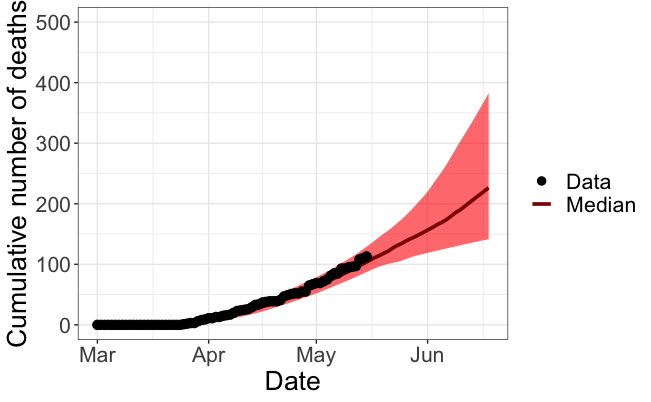}\\
    \includegraphics[scale=0.25]{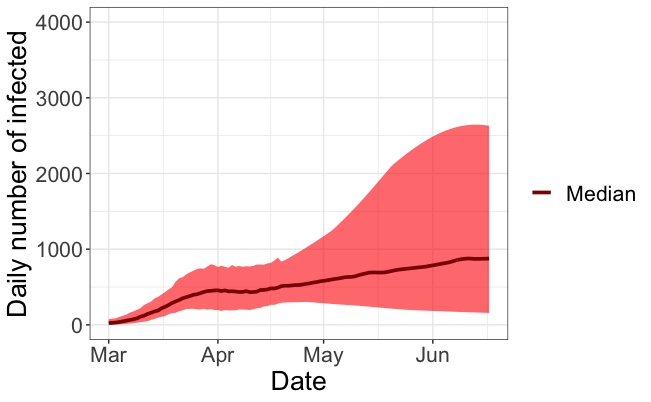}
    \includegraphics[scale=0.25]{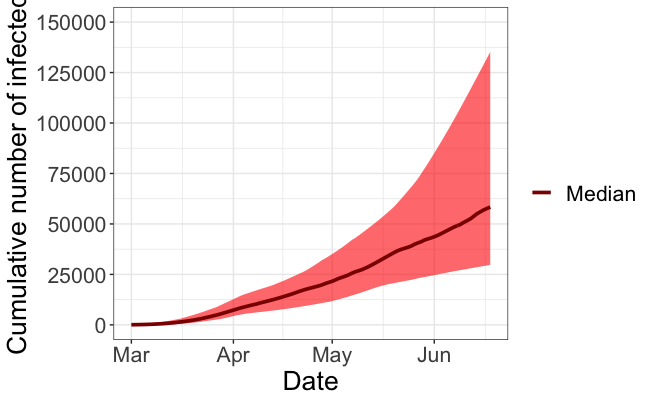}
    \caption{Data from Region J\"onk\"oping  until 2020-05-15. \emph{Top left:} Observed daily number of deaths (black dots), the posterior median (dark red) and $95\%$ credibility interval for the expected daily number of deaths. \emph{Top right:} Observed cumulative number of deaths (black dots), the posterior median (dark red) and $95\%$ credibility interval for the expected cumulative number of deaths. \emph{Bottom left:} The posterior median (dark red) and $95\%$ credibility interval for the daily number of infected individuals. \emph{Bottom right:} The posterior median (dark red) and $95\%$ credibility interval for the cumulative number of infected individuals. }
    \label{fig:Daily_deaths_JON}
\end{figure}

\begin{figure}
    \centering
    \includegraphics[scale=0.25]{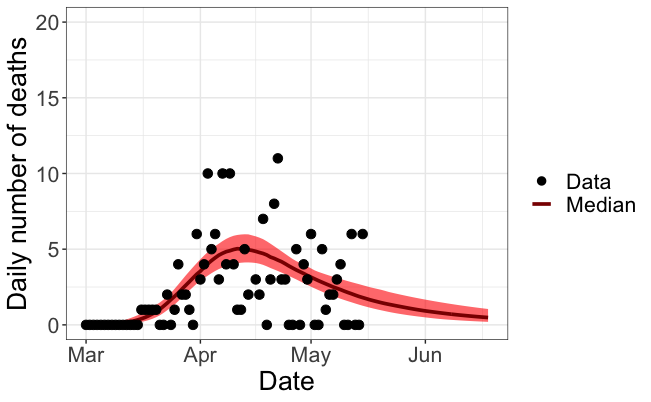}
    \includegraphics[scale=0.25]{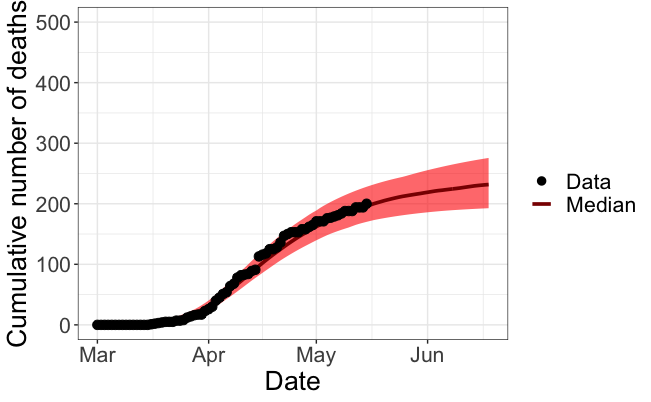}\\
    \includegraphics[scale=0.25]{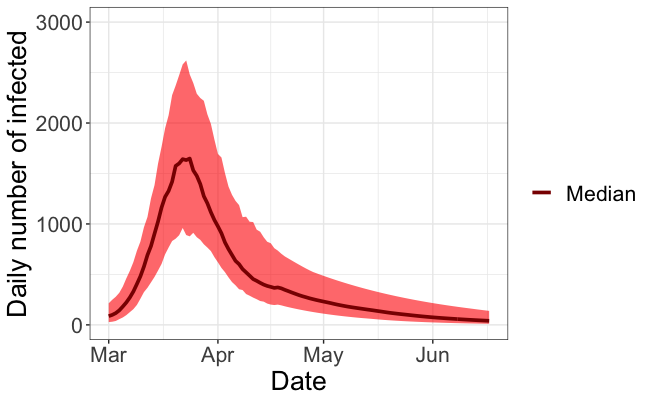}
    \includegraphics[scale=0.25]{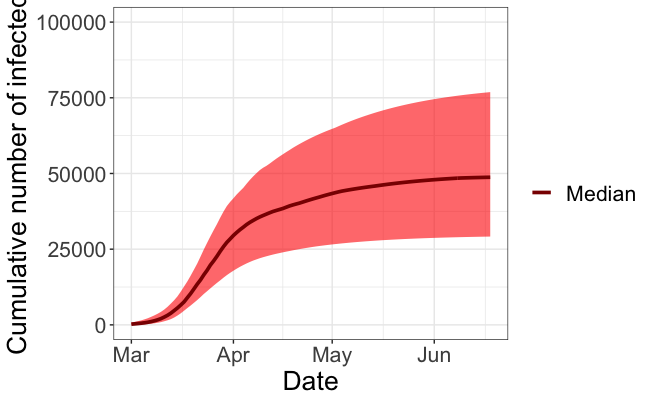}
    \caption{Data from Region S\"ormland  until 2020-05-15. \emph{Top left:} Observed daily number of deaths (black dots), the posterior median (dark red) and $95\%$ credibility interval for the expected daily number of deaths. \emph{Top right:} Observed cumulative number of deaths (black dots), the posterior median (dark red) and $95\%$ credibility interval for the expected cumulative number of deaths. \emph{Bottom left:} The posterior median (dark red) and $95\%$ credibility interval for the daily number of infected individuals. \emph{Bottom right:} The posterior median (dark red) and $95\%$ credibility interval for the cumulative number of infected individuals. }
    \label{fig:Daily_deaths_SOR}
\end{figure}

\begin{figure}
    \centering
    \includegraphics[scale=0.25]{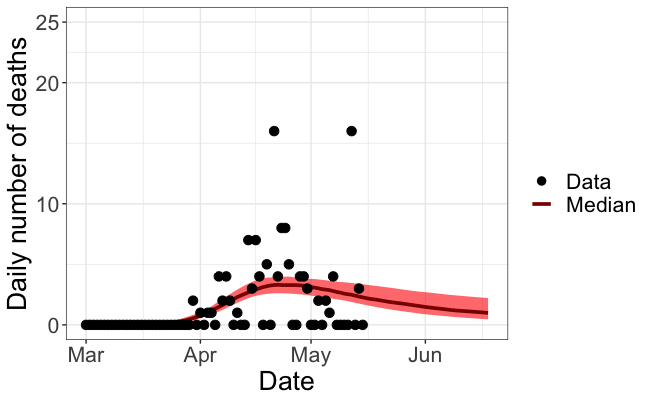}
    \includegraphics[scale=0.25]{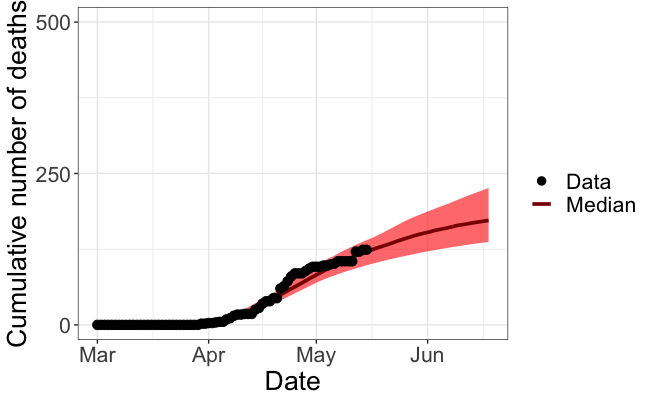}\\
    \includegraphics[scale=0.25]{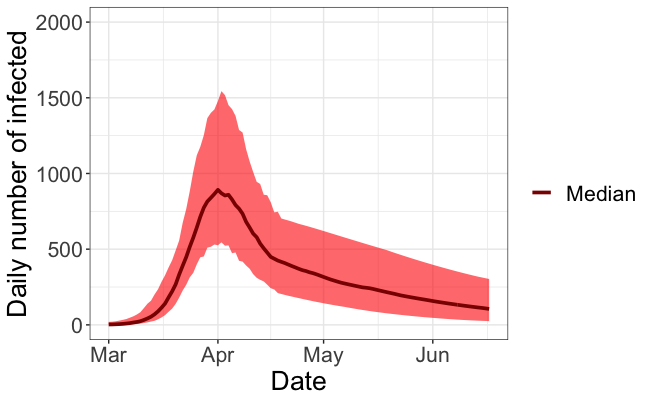}
    \includegraphics[scale=0.25]{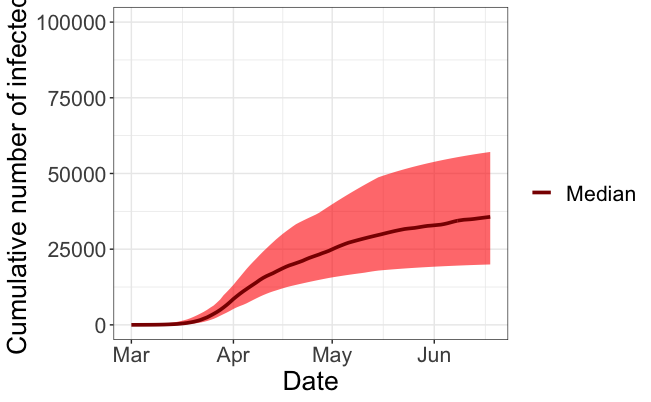}
    \caption{Data from Region V\"astmanland  until 2020-05-15. \emph{Top left:} Observed daily number of deaths (black dots), the posterior median (dark red) and $95\%$ credibility interval for the expected daily number of deaths. \emph{Top right:} Observed cumulative number of deaths (black dots), the posterior median (dark red) and $95\%$ credibility interval for the expected cumulative number of deaths. \emph{Bottom left:} The posterior median (dark red) and $95\%$ credibility interval for the daily number of infected individuals. \emph{Bottom right:} The posterior median (dark red) and $95\%$ credibility interval for the cumulative number of infected individuals. }
    \label{fig:Daily_deaths_VST}
\end{figure}

\begin{figure}
    \centering
    \includegraphics[scale=0.25]{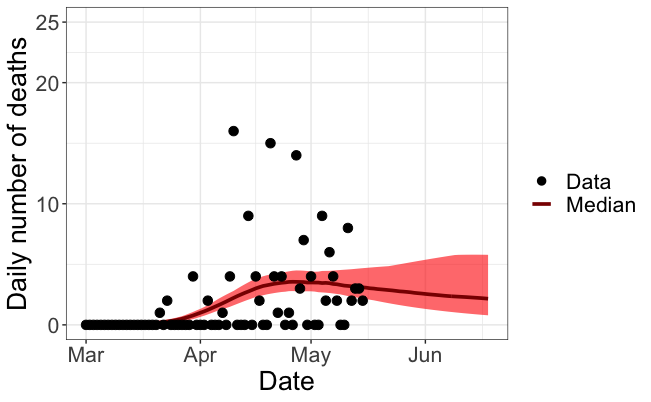}
    \includegraphics[scale=0.25]{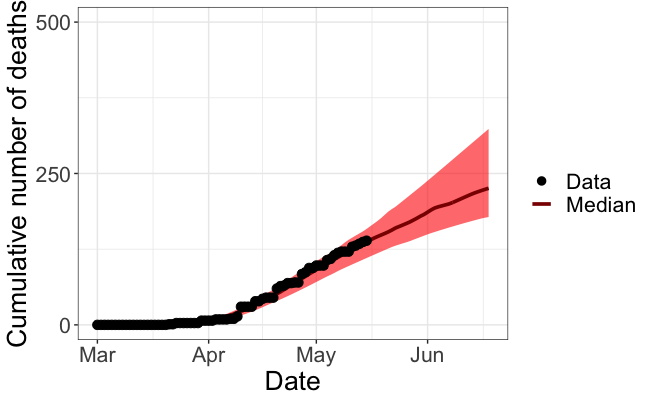}\\
    \includegraphics[scale=0.25]{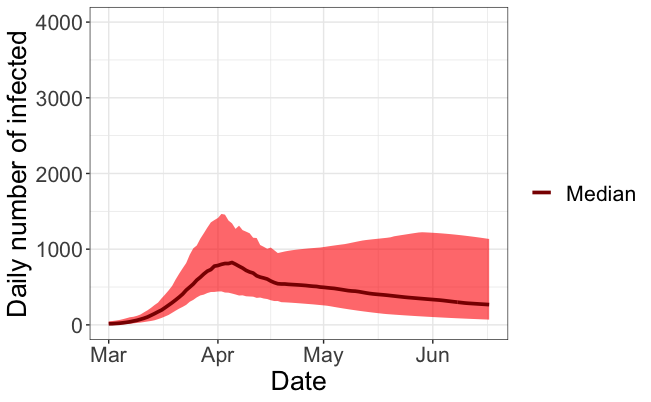}
    \includegraphics[scale=0.25]{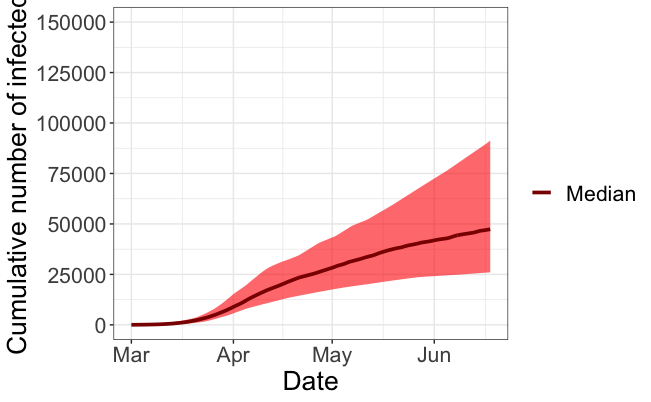}
    \caption{Data from Region Uppsala  until 2020-05-15. \emph{Top left:} Observed daily number of deaths (black dots), the posterior median (dark red) and $95\%$ credibility interval for the expected daily number of deaths. \emph{Top right:} Observed cumulative number of deaths (black dots), the posterior median (dark red) and $95\%$ credibility interval for the expected cumulative number of deaths. \emph{Bottom left:} The posterior median (dark red) and $95\%$ credibility interval for the daily number of infected individuals. \emph{Bottom right:} The posterior median (dark red) and $95\%$ credibility interval for the cumulative number of infected individuals. }
    \label{fig:Daily_deaths_UPP}
\end{figure}

\begin{figure}
    \centering
    \includegraphics[scale=0.25]{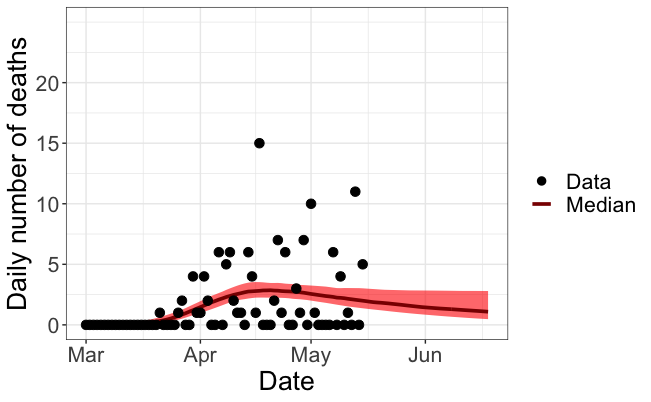}
    \includegraphics[scale=0.25]{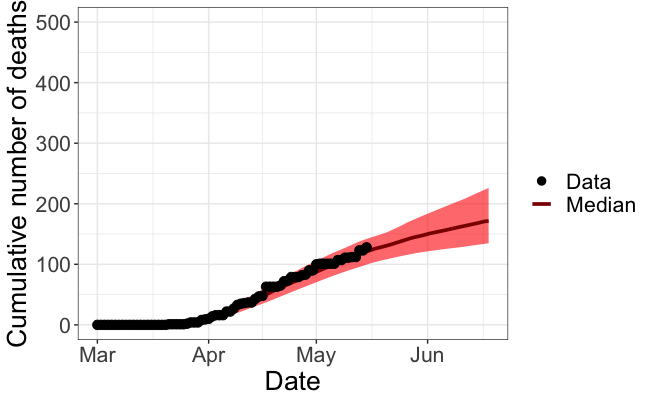}\\
    \includegraphics[scale=0.25]{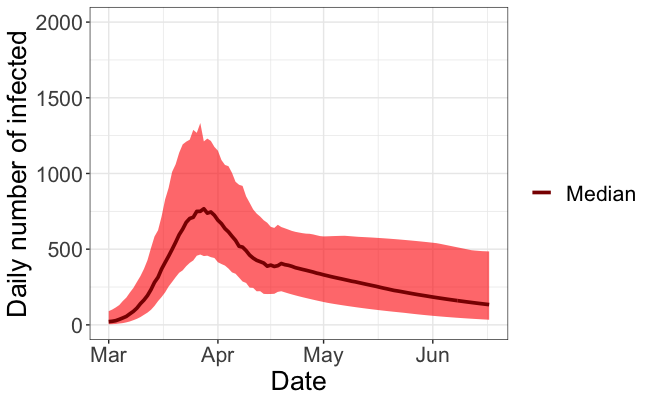}
    \includegraphics[scale=0.25]{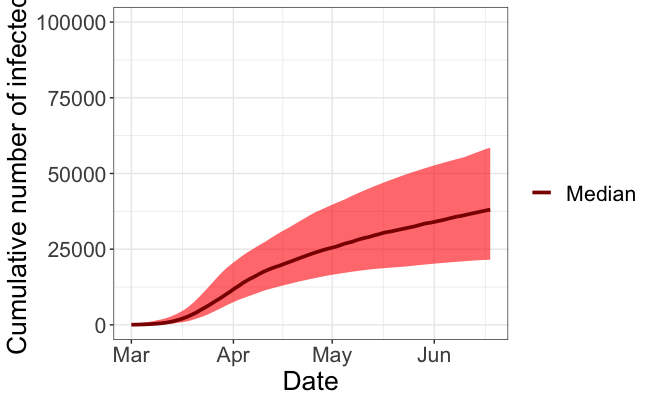}
    \caption{Data from Region Dalarna  until 2020-05-15. \emph{Top left:} Observed daily number of deaths (black dots), the posterior median (dark red) and $95\%$ credibility interval for the expected daily number of deaths. \emph{Top right:} Observed cumulative number of deaths (black dots), the posterior median (dark red) and $95\%$ credibility interval for the expected cumulative number of deaths. \emph{Bottom left:} The posterior median (dark red) and $95\%$ credibility interval for the daily number of infected individuals. \emph{Bottom right:} The posterior median (dark red) and $95\%$ credibility interval for the cumulative number of infected individuals. }
    \label{fig:Daily_deaths_DAL}
\end{figure}

The daily death counts for the regions of Sweden until 2020-05-15 are obtained from the webpage: \verb=https://c19.se/=. There is no random testing providing information on the proportion of infected individuals outside Region Stockholm. To estimate the contact rate and the cumulative number of infected individuals in regions outside Stockholm, we have implemented the multi-region model pairwise, with two regions in each MCMC simulation, where one region is  Stockholm and the other region is from the list above. It is assumed that the parameters $\nu, \gamma, f,$ and $s_D$ are identical in both regions, but the time varying contact rate and the initial proportion of susceptible individuals are different between the regions. The posterior of the contact rates for the different regions are provided in Figure \ref{fig:Contact_rates} and the corresponding estimates and predictions for the daily number of deaths, the cumulative number of deaths, the daily number of infected individuals and the cumulative number of infected individuals are provided in Figures \ref{fig:Daily_deaths_VGO} - \ref{fig:Daily_deaths_DAL}. The parameter uncertainty is generally high and in several regions the number of new infections and daily deaths may still increase. Estimates on the proportion of infected individuals on 2020-05-15 in the different regions are provided in Table \ref{tab:proportions} along with $95\%$ credibility intervals. Overall the proportions are low, far from herd immunity.

\begin{table}[!ht]
    \centering
    \begin{tabular}{l|r|r}
        Region &  Prop. infected & $95\%$-C.I. \\
        \hline
        Stockholm & $13.5\%$  & $[8.4\%, 20.1\%]$\\
        V\"astra G\"otaland &$5.3\%$ & $[3.1 \%,9.5 \%]$\\
        \"Osterg\"otland & $8.3\%$& $[4.8\%, 13.8\%]$\\
        \"Orebro &$9.9\%$ & $[5.8\%, 17.0\%]$\\
        Sk{\aa}ne & $2.5\%$  & $[1.2\%, 4.5\%]$\\
        J\"onk\"oping & $8.8\%$ & $[ 5.2\%, 14.5\%]$\\
        S\"ormland & $15.6\%$ & $[9.45\%, 23.9\%]$\\
        V\"astmanland &$10.8\%$ & $[6.5 \%, 17.8\%]$\\
        Uppsala &$9.4\%$ & $[5.6\%, 14.7\%]$\\
        Dalarna &$ 10.4\%$ & $[6.5 \%, 16.1\%]$\\
         \hline
    \end{tabular}
    \caption{Estimated proportion of the population who have had a SARS-CoV-2 infection by May 15, for ten regions of Sweden. }
    \label{tab:proportions}
\end{table}



\begin{thebibliography}{99}
\bibitem{Anderson92}
R.M.\ Anderson and R.M.\ May. 
\newblock  \emph{Infectious Diseases of Humans: Dynamics and Control}.
\newblock Oxford University Press, 1992.
\bibitem{Betancourt17}
M.\ Betancourt, S.\ Byrne, S.\ Livingstone, and M.\ Girolami.
\newblock The geometric
foundations of Hamiltonian Monte Carlo. 
\newblock \emph{Bernoulli} 23 (4A), 2257–2298, 2017. 
\bibitem{Britton20}
T.~Britton. 
\newblock Basic estimation-prediction techniques for COVID-19,
and a prediction for Stockholm. 
\newblock \emph{MedRxiv} https://doi.org/10.1101/2020.04.15.20066050
\bibitem{Chatzilena19}
A.~Chatzilena, E.~van Leeuwen, O.~Ratmann, M.~Baguelin, and N. Demiris.
\newblock Contemporary statistical inference for infectious disease models using Stan.
\newblock \emph{Epidemics}, 29, 100367, 2019.
\bibitem{Diekmann13}
O.\ Diekmann, J.A.P.\ Heesterbeek, and T.\ Britton.
\newblock \emph{Mathematical tools for
understanding infectious disease dynamics.}
\newblock Princeton University Press, 2013. 
\bibitem{Geman84}
S.\ Geman, and D.\ Geman, D.
\newblock Stochastic relaxation, Gibbs distributions, and the Bayesian
restoration of images. 
\newblock \emph{IEEE Trans. Pattern Anal. Mach. Intell.} (6), 721–741, 1984.
\bibitem{Griewank08}
A.\ Griewank, and A.\  Walther, A.
\newblock \emph{Evaluating Derivatives: Principles and Techniques of
Algorithmic Differentiation}, SIAM, 2008.
\bibitem{Hastings70}
W.K.\ Hastings. 
\newblock Monte Carlo sampling methods using Markov Chains and their
applications, 
\newblock \emph{Biometrika}, 57(1), 97–109,1970.
\bibitem{Hoffmann14}
M.D.\ Hoffmann and A.\ Gelman. 
\newblock The No-U-turn sampler: adaptively setting path lengths
in Hamiltonian Monte Carlo. 
\newblock \emph{J. Mach. Learn. Res.} 15 (1), 1593–1623, 2014.
\bibitem{Kallenberg17}
O.\ Kallenberg.
\newblock \emph{Random Measures, Theory and Applications}. 
\newblock Springer, 2017.
\bibitem{LiGuan20}
Q.\ Li et al.  
\newblock Early transmission dynamics in Wuhan, China, of novel Coronavirus 
\newblock \emph{Infected Pneumonia. N. Engl. J. Med.} 382, 1199-1207, 2020.
\bibitem{LiPei20} 
Li, R., Pei, S., Chen, B., Song, Y., Zhang, T., Yang, W. and Shaman, J. (2020).
\newblock Substantial undocumented infection facilitates the rapid dissemination of novel coronavirus (SARS-CoV2). 
\newblock \emph{Science}, 10.1126/science.abb3221
\bibitem{Metropolis53}
N.\ Metropolis, A.W.\ Rosenbluth, M.N.\ Rosenbluth, A.H.\ Teller, E.\ Teller.
\newblock Equation of state calculations by fast computing machines. 
\newblock \emph{J. Chem. Phys.} 21 (6), 1087–1092, 1953.
\bibitem{Verity20}
R. Verity et al.
\newblock Estimates of the severity of coronavirus disease 2019:
a model-based analysis
\newblock \emph{The Lancet Infections Diseases}, March 30, 2020, 
\bibitem{Wu20}
J.T.\ Wu et al. 
\newblock Estimating clinical severity of COVID-19 from the transmission dynamics in Wuhan, China. 
\newblock \emph{Nature Medicine} 26, 506–510, 2020. 
\end{thebibliography}
\end{document}